\documentclass[10pt,aps,prb,twocolumn,superscriptaddress]{revtex4-2}
\usepackage[utf8]{inputenc}
\usepackage{graphics}
\usepackage{epsfig}
\usepackage{dcolumn}
\usepackage{bm}
\usepackage{amsmath}
\usepackage{dsfont} 
 \usepackage{mathtools}
 \usepackage{amsfonts}
\usepackage{latexsym}
\usepackage{amssymb}
\usepackage{color}
\usepackage{hyperref}

\hypersetup{
    colorlinks=true,
    linkcolor=blue,
    filecolor=magenta,      
    urlcolor=cyan,
}
\usepackage[normalem]{ulem}
 \usepackage{bbm}
\usepackage{bbold}

\usepackage{amsfonts}

 \newcommand{\g}{\bf}
\usepackage{braket}

\begin{document}
\title{Gutzwiller approximation for paramagnetic ionic Hubbard model:\\
Analytic expression for  band - Mott insulator transition}
\thanks{This work is dedicated to Professor Józef Spałek on the occasion of his 80th birthday.}
 \author{Marcin M. Wysoki\'nski}
 \email{wysokinski@magtop.ifpan.edu.pl}

  \affiliation{International Research Centre MagTop, Institute of
   Physics, Polish Academy of Sciences,\\ Aleja Lotnik\'ow 32/46,
   PL-02668 Warsaw, Poland}

\begin{abstract}

The ionic Hubbard model is a paradigmatic setup for studying the competition between band and Mott insulating behavior. Within the variationally exact in infinite dimensions Gutzwiller approximation, we derive a compact analytic expression for the phase boundary between Mott and band insulator. While the method reproduces the expected band–Mott insulator phenomenology, it does not capture the correlated metallic state at finite staggered potential found for example in dynamical mean-field theory. This absence highlights that the metallic phase originates from incoherent Hubbard-band physics rather than Fermi-liquid behavior well captured by Gutzwiller approximation. Our formulation establishes a concise variational framework to ionic Hubbard model, with natural extensions to nonequilibrium setups and spin-exchange dynamics.
\end{abstract}  
\maketitle

\section{Introduction}

The Hubbard model is probably the simplest yet remarkably rich quantum mechanical platform for describing correlated electrons on a crystallographic lattice. Its continuous appeal within the condensed matter community stems among others from its ability to capture essential features of Mott physics.

At half-filling, the model hosts a Mott insulating state characterized by frozen charge degrees of freedom due to strong onsite Coulomb repulsion, while spin degrees of freedom remain active via antiferromagnetic exchange. As shown by Spałek in his derivation of the $t$-$J$ model \cite{Spalek_1977} and in later analyses, spin dynamics remain active even away from half-filling. The nontrivial role of correlations beyond the insulating regime, as described by the $t$-$J$ model, is revealed by the emergence of unconventional superconductivity \cite{Jedrak,Zegrodnik2,Zegrodnik1}.

The Hubbard model is also very versatile, as seemingly small extensions to its pristine form allow us to analyze a quantum-mechanical description of various phenomena on the verge of Mott physics. This is the case for the Hubbard model enriched by a staggered potential on a bipartite lattice (i.e., an ionic potential) that captures the band-to-Mott insulator transition \cite{Fabrizio_1999, Torio_2001,Painelli_2001,Martin_2001, Kampf_2003, Manmana_2004,Jabben_2005,Garg_2006, Scalettar_2007, Dagotto_2007,Craco_2008, Byczuk_2009,Baeriswyl_2009,Hoang_2010}, and thus in the past has been proposed to capture the ionic-to-neutral insulator transition in organic charge-transfer solids \cite{LaPlaca_1981}, and ferroelectric transitions in perovskites \cite{Egami_1993}.

More recently, the model has been studied in the context of flat bands \cite{Rubio_2020}, cold-atom setups \cite{Esslinger_2015,Kollath_2017,Esslinger_2024}, topological properties of the dimerized (bond-ordered) phase (initially discussed in Ref. \onlinecite{Fabrizio_1999}) that may occur between the BI/MI phase boundary \cite{Aligia_2023}, and within the context of time-dependent phenomena \cite{Kollath_2017,Kollath_2022}. 

Despite this broad interest, applications of the Gutzwiller approximation (GA), a popular mean-field many-body tool, to the ionic Hubbard model remain limited \cite{Garg_2019,Wang_2022}. Here we develop a neat formulation of the Gutzwiller approximation based on a general projector \cite{Lanata2008,Fabrizio2007,Fabrizio2012}. This formulation is ready to be extended to the non-equilibrium set-up \cite{Schiro2010,Fabrizio2012}, as well as to include spin-exchange physics in a variational manner \cite{Wysokinski2016,Wysokinski2017}. In this work, however, we restrict our analysis to the equilibrium and low-energy physics of charge degrees of freedom in the paramagnetic ionic Hubbard model.

With the use of GA, we have derived an analytic formula for the band–Mott insulator phase boundary. In contrast to dynamical mean-field theory analyses, which reveal a metallic phase at the band–Mott insulator boundary in infinite dimensions \cite{Garg_2006,Byczuk_2009}, our approach does not reproduce this feature. We attribute its absence to the fact that the GA inherently describes only Fermi-liquid behavior, thereby missing incoherent effects responsible for the metallic state.

\section{Model}
The half-filled ionic Hubbard model describes electrons moving on a bipartite lattice under the influence of both a staggered potential and onsite Coulomb interactions. Its Hamiltonian is
\begin{equation}
    \begin{split}
    \mathcal{H}&=-t\sum_{\langle{\g i}{\g j}\rangle\sigma}c_{\g i\sigma}^\dagger c_{\g j\sigma}+\sum_{\g i}\Big(\frac{U}{2} (n_{{\g i}} -1)^2+\Delta(-1)^{\g i}n_{{\g i}}\Big) 
    \label{model}
    \end{split}
\end{equation}
where $  n_{\g i}\equiv n_{{\g i}\uparrow }+n_{{\g i}\downarrow }$,  and $(-1)^{{\g i}\in A}=1$  $(-1)^{{\g i}\in B}=-1$ distinguishes the two sublattices. In this formulation, 
$t>0$ represents the hopping integral between nearest neighbors, 
$U$ is the onsite Coulomb repulsion, and $\Delta$ sets the amplitude of the staggered ionic potential alternating between the two sublattices. The model possesses a combined particle–hole and sublattice-exchange symmetry (A$\to$B), which enforces half-filling on average. Consequently, there is no explicit chemical potential term.

\section{Gutzwiller Approximation}
In order to analyze the properties of the model (\ref{model}), we use an efficient formulation of the Gutzwiller approximation \cite{Lanata2008,Fabrizio2007,Fabrizio2012}, in which we define a general local Gutzwiller projector as
\begin{equation}
 \mathcal{P}_{\g i}=\sum_\Gamma\frac{\Phi_{{\g i}\Gamma}}{\sqrt{\prod_\sigma (n^0_{{\g i}\sigma }) ^{n_\sigma}(1-n^0_{{\g i}\sigma})^{(1-n_\sigma)}}}|{\g i} ;\Gamma\rangle\langle{\g i};\Gamma|  
\end{equation}
where local Fock states are defined as ${|{\g i};\Gamma\rangle=\prod_\sigma(c_{\g i,\sigma}^\dagger)^{n_\sigma}|0\rangle}$ i.e. ${|{\g i};\Gamma\rangle\in\{|0\rangle,|\uparrow\downarrow\rangle,|\uparrow\rangle,|\downarrow\rangle\}}$. The Gutzwiller approximation, variationally exact only in the limit of infinite dimensions, is enforced by imposing the following constraints:  
  \begin{equation}
  \begin{split}
  &{\rm Tr}(\hat \Phi_{\g i}^\dagger\hat \Phi_{\g i})=1\\
  &{\rm Tr}(\hat \Phi_{\g i}^\dagger\hat \Phi_{\g i}\hat c_{{\g i} \sigma}^\dagger\hat c_{{\g i} \sigma'})=n_{{\g i}\sigma}^0 \delta_{\sigma,\sigma'}\\
  \label{GA}
  \end{split}
  \end{equation}
  where hat denotes the matrix representation of the object in the $|{\g i};\Gamma\rangle$ base.
  Conveniently, the expectation value under the Gutzwiller approximation of any local operator can be calculated as $\langle \mathcal O_{\g i}\rangle={\rm Tr}[\hat\Phi^\dagger_{\g i}\mathcal{\hat O}_{\g i}\Phi_{\g i}]$. 
While implementing the first constraint is relatively straightforward, the second one needs to be treated with some care, and specifically, we enforce it here through Lagrange multipliers 
$\lambda_i$. Eventually, we can define an effective single-particle Hamiltonian as
\begin{equation}
 \begin{split}
 \mathcal{H}_* =-t&\sum_{\langle{\g i}{\g j}\rangle\sigma}(R^*_{i\sigma}R_{j\sigma}  c_{\g i\sigma}^\dagger  c_{\g j\sigma}+{\rm H.c.})\\
 + \!\sum_{\g i}&\!\Bigg[\!\frac{U}{2}{\rm Tr}\big[\hat \Phi_{\g i}^\dagger(\hat n_{\g i}\!-\!1)^2\hat \Phi_{\g i}\big]\!+\!\Delta (\!-1)^{\g i}{\rm Tr}\big[\hat \Phi_{\g i}^\dagger \hat n_{\g i}\hat \Phi_{\g i}\big]  \\
 &+ \lambda_{\g i}\Big( n_{\g i}-{\rm Tr}\big[\hat \Phi_{\g i}^\dagger \hat n_{\g i}\hat \Phi_{\g i}\big]\Big)\Bigg]\\
 \end{split}
 \end{equation}
satisfying $\langle \mathcal{H}\rangle=\langle \mathcal{H}_*\rangle_0$ ($\langle ...\rangle_0$ denotes an expectation value with respect to the Slater determinant), and where 
 \begin{equation}
 \begin{split}
     &R^*_{{\g i}\sigma}=\frac{1}{\sqrt{n^0_{{\g i}\sigma}(1-n^0_{{\g i}\sigma})}}{\rm Tr}(\hat \Phi_{\g i}^\dagger \hat c_{{\g i} \sigma}^\dagger\hat \Phi_{\g i} \hat c_{{\g i} \sigma})
 \end{split}
 \end{equation}

Given a particle-hole with an 
$A\to B$
 shift symmetry of our Hamiltonian, and our interest in the paramagnetic state, we can write
\begin{equation}
    \begin{split}
    \hat \Phi_{{\g i}\in A}&=\frac{1}{\sqrt{2}}\begin{pmatrix}
    \phi_0&0&0&0\\
    0&\phi_2&0&0\\
    0&0&\phi_1&0\\
    0&0&0&\phi_1
    \end{pmatrix}\\
    \hat \Phi_{{\g i}\in B}&=\frac{1}{\sqrt{2}}\begin{pmatrix}
    \phi_2&0&0&0\\
    0&\phi_0&0&0\\
    0&0&\phi_1&0\\
    0&0&0&\phi_1
    \end{pmatrix}
    \end{split}
\end{equation} 
and $\lambda_{{\g i}\in A}=-\lambda_{{\g i}\in B}=\lambda$.

  At equilibrium all $\phi_\alpha$ parameters can be real, and thus we can make a further parameterization
  \begin{equation}
      \begin{split}
      \phi_0&=\sqrt{2}\sin\theta \cos\varphi\\
      \phi_2&=\sqrt{2}\sin\theta \sin\varphi\\
      \phi_1&=\cos\theta
      \end{split}
  \end{equation}
   where it is enough to assume $\{\theta,\varphi\}\in\{0,\pi/2\}$. With the above, we automatically satisfy the first Gutzwiller approximation constraint (cf. \eqref{GA}). Moreover, we have
   \begin{equation}
     \begin{split}
  &n_A={\rm Tr}(\hat \Phi_{\g i}^\dagger\hat \Phi_{\g i} \hat n_{{\g i}\in A })=|\phi_1|^2+|\phi_2|^2 \\
  &n_B={\rm Tr}(\hat \Phi_{\g i}^\dagger\hat \Phi_{\g i} \hat n_{{\g i}\in B })=|\phi_1|^2+|\phi_0|^2.\\
  &R_A=R_{{\g i}\in A\sigma}=\frac{1}{\sqrt{n_An_B }}(\phi_0^*\phi_1+\phi_1^*\phi_2)=R_B^*
  \end{split}
   \end{equation}
In the above, for the sake of clarity, we leave complex conjugations even if 
$\phi_\alpha$
 are real in equilibrium. 
 Finally, the variational energy per lattice site, $\langle \mathcal{H}_*\rangle_0/N$ reads
\begin{equation}
    \begin{split}
E(\lambda,\varphi,\theta)&=T(\lambda,\varphi,\theta)+\sin^2\theta\Big(\frac{U}{2}+\cos2\varphi  (\lambda-\Delta) \Big)
    \end{split}
\end{equation}
with $T(\lambda,\varphi,\theta)$ being the energy of the single-particle effective Hamiltonian. 
We assume a simple rectangular density of states $N(\epsilon)=1/2W$, where 
$W$ is half of the bandwidth, and a dispersion that has the property $\epsilon_{k+Q}=-\epsilon_k$, so that
\begin{equation}
   T(\lambda,\varphi,\theta)= -\frac{1}{2W}\int_{-W}^{W}\!\!\!d\epsilon \, \sqrt{R^4\epsilon^2+\lambda^2}
\end{equation}
where spin degeneracy is compensated by the integral over the whole Brillouin zone. The saddle point of the energy functional can be found by solving a set of differential equations
\begin{equation}
    \{\frac{\partial}{\partial\lambda},\frac{\partial}{\partial\varphi},\frac{\partial}{\partial\theta}\}E=0.
    \label{set}
\end{equation}
In that manner, the saddle point solution can differentiate between three phases:
\begin{itemize}
    \item band insulator ($\lambda>0$ and $\theta>0$)
    \item correlated metal ($\lambda=0$ and $\theta>0$)
    \item Mott insulator ($\theta=0$).
\end{itemize}

The first issue that we can immediately resolve is that equations \eqref{set} cannot lead to the correlated metal state for non-zero $\Delta$. At  $\lambda=0$, equations \eqref{set} do not lead to a contradiction only if $\theta=0$. For that reason, apart from the limiting $U=0$
 case, our approach leads only to band or Mott insulating phases on the $\Delta$-$U$ plane, separated by the Brinkman-Rice (BR) transition line.

\begin{figure}[t!]
    \centering
    \includegraphics[width=0.5\textwidth]{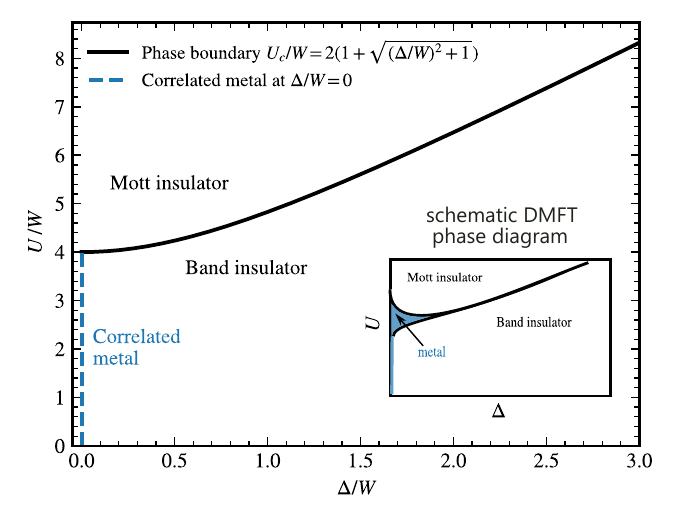}
    \caption{Phase diagram in the $(\Delta/W, U_c/W)$ plane. The solid line denotes the phase boundary between the Mott insulating and band insulating states, given by $U_c/W = 2\left(1+\sqrt{(\Delta/W)^2+1}\right)$. At $\Delta/W=0$, the system exhibits a correlated metallic phase up to $U_c/W=4$. The region above the boundary corresponds to the Mott insulator, while the region below corresponds to the band insulator. In the inset we provide a schematic phase diagram obtained from dynamical mean-field theory after Ref.~\onlinecite{Garg_2006}.}
    \label{fig1}
\end{figure}

We solve the equations \eqref{set} numerically and observe that near the BR transition, $\lambda$ goes to zero much faster than $\theta$. Due to the apparent singularity in the kinetic energy at the transition, in the following we determine the transition line analytically. We do this by analyzing the equations \eqref{set} for $\lambda=0$ in linear order of small $\theta$
\begin{equation}
\begin{split}
    &\frac{\partial E}{\partial \theta}|_{\lambda=0}=\theta\big[U-2(W+\Delta\cos 2\varphi+W\sin 2\varphi)\big]\\
    &\frac{\partial E}{\partial \varphi}|_{\lambda=0}=\frac{\theta}{2}\big[\Delta\sin2\varphi-W\cos2\varphi \big]
\end{split}
\end{equation}
From the above it follows that near the transition
\begin{equation}
\cot2\varphi=\Delta/W
\label{cot}
\end{equation} 
and therefore the critical interaction strength $U_c$ for which the BR transition takes place is determined by 
\begin{equation}
  U_c/W=2(1+\sqrt{(\Delta/W)^2+1})  
\end{equation} 
If $\Delta=0$ it follows (cf. Eq. (\ref{cot})) that $\varphi=\pi/4$  and  we restore the usual relation $U_c/W=4$ for the BR transition in the Hubbard model.

 \section{Discussion}
The Gutzwiller approximation is a many-body approach that becomes variationally exact in the limit of infinite dimensions. Although it does not capture high-energy quantum fluctuations responsible for spin exchange \cite{Wysokinski2016}, it provides a good description of Fermi-liquid behavior. This is achieved primarily through its treatment of quasiparticle renormalization and the shift in chemical potential, both arising from the low-energy expansion of the real part of the self-energy. Moreover, the method successfully captures the freezing of charge degrees of freedom with increasing correlations, signaling the metal–Mott insulator transition deemed Brinkman-Rice transition.

As expected, our approach reproduces the phenomenology of the band-to-Mott insulator transition on the $\Delta$-$U$ plane, well established for the infinitely dimensional Hubbard model \cite{Garg_2006}. However, given its ability to describe Fermi-liquid behavior, it is somewhat surprising that, by construction, our approach does not yield a correlated metallic state for nonzero $\Delta$ in the vicinity of band to Mott insulator transition \cite{Garg_2006,Byczuk_2009}.

This observation, however, further illustrates \cite{Hoang_2010} that the origin of the metallic phase is not of the Fermi-liquid type. Instead, it is rooted in incoherent features and interplay between Hubbard and ionic bands splits, that near the transition result in the non-zero spectral function at the Fermi level.
 
 \section{Summary}
 In this work we have analyzed the half-filled ionic Hubbard model within the Gutzwiller approximation, focusing on the competition between band and Mott insulating phases. By constructing a compact variational formulation based on general local projectors, we derived an analytic expression for the phase transition line separating the two insulating regimes. Our analysis shows that, while the Gutzwiller method correctly reproduces the expected band–Mott insulator phenomenology, it does not support the existence of a correlated metallic phase at finite staggered potential that has been observed via dynamical mean-field theory \cite{Garg_2006,Byczuk_2009}. Given intrinsic Fermi-liquid character of the GA, neglecting incoherent features related to Mott physics we attribute emergence of metallic phase in the vicinity of the phase boundary between different insulators to the presence of spectral features of Mott-Hubbard bands. Our results, in broad picture, highlight the complementary roles of different many-body approaches. 
Moreover, the formulation itself is readily extendable to nonequilibrium setups \cite{Schiro2010,Fabrizio2012} and to variational treatments of spin-exchange physics \cite{Wysokinski2016,Wysokinski2017}.  
\section*{Acknowledgments}
This research was supported by the “MagTop” project (FENG.02.01-IP.05-0028/23) carried out within the “International Research Agendas” programme of the Foundation for Polish Science co-financed by the European Union under the European Funds for Smart Economy 2021-2027 (FENG).


\begin{thebibliography}{34}%
\makeatletter
\providecommand \@ifxundefined [1]{%
 \@ifx{#1\undefined}
}%
\providecommand \@ifnum [1]{%
 \ifnum #1\expandafter \@firstoftwo
 \else \expandafter \@secondoftwo
 \fi
}%
\providecommand \@ifx [1]{%
 \ifx #1\expandafter \@firstoftwo
 \else \expandafter \@secondoftwo
 \fi
}%
\providecommand \natexlab [1]{#1}%
\providecommand \enquote  [1]{``#1''}%
\providecommand \bibnamefont  [1]{#1}%
\providecommand \bibfnamefont [1]{#1}%
\providecommand \citenamefont [1]{#1}%
\providecommand \href@noop [0]{\@secondoftwo}%
\providecommand \href [0]{\begingroup \@sanitize@url \@href}%
\providecommand \@href[1]{\@@startlink{#1}\@@href}%
\providecommand \@@href[1]{\endgroup#1\@@endlink}%
\providecommand \@sanitize@url [0]{\catcode `\\12\catcode `\$12\catcode
  `\&12\catcode `\#12\catcode `\^12\catcode `\_12\catcode `\%12\relax}%
\providecommand \@@startlink[1]{}%
\providecommand \@@endlink[0]{}%
\providecommand \url  [0]{\begingroup\@sanitize@url \@url }%
\providecommand \@url [1]{\endgroup\@href {#1}{\urlprefix }}%
\providecommand \urlprefix  [0]{URL }%
\providecommand \Eprint [0]{\href }%
\providecommand \doibase [0]{https://doi.org/}%
\providecommand \selectlanguage [0]{\@gobble}%
\providecommand \bibinfo  [0]{\@secondoftwo}%
\providecommand \bibfield  [0]{\@secondoftwo}%
\providecommand \translation [1]{[#1]}%
\providecommand \BibitemOpen [0]{}%
\providecommand \bibitemStop [0]{}%
\providecommand \bibitemNoStop [0]{.\EOS\space}%
\providecommand \EOS [0]{\spacefactor3000\relax}%
\providecommand \BibitemShut  [1]{\csname bibitem#1\endcsname}%
\let\auto@bib@innerbib\@empty
\bibitem [{\citenamefont {Chao}\ \emph {et~al.}(1977)\citenamefont {Chao},
  \citenamefont {Spalek},\ and\ \citenamefont {Oles}}]{Spalek_1977}%
  \BibitemOpen
  \bibfield  {author} {\bibinfo {author} {\bibfnamefont {K.~A.}\ \bibnamefont
  {Chao}}, \bibinfo {author} {\bibfnamefont {J.}~\bibnamefont {Spalek}},\ and\
  \bibinfo {author} {\bibfnamefont {A.~M.}\ \bibnamefont {Oles}},\ }\bibfield
  {title} {\bibinfo {title} {Kinetic exchange interaction in a narrow s-band},\
  }\href {https://doi.org/10.1088/0022-3719/10/10/002} {\bibfield  {journal}
  {\bibinfo  {journal} {Journal of Physics C: Solid State Physics}\ }\textbf
  {\bibinfo {volume} {10}},\ \bibinfo {pages} {L271} (\bibinfo {year}
  {1977})}\BibitemShut {NoStop}%
\bibitem [{\citenamefont {Jedrak}\ and\ \citenamefont
  {Spa\l{}ek}(2011)}]{Jedrak}%
  \BibitemOpen
  \bibfield  {author} {\bibinfo {author} {\bibfnamefont {J.}~\bibnamefont
  {Jedrak}}\ and\ \bibinfo {author} {\bibfnamefont {J.}~\bibnamefont
  {Spa\l{}ek}},\ }\bibfield  {title} {\bibinfo {title} {Renormalized mean-field
  $t\ensuremath{-}j$ model of high-${T}_{c}$ superconductivity: Comparison to
  experiment},\ }\href {https://doi.org/10.1103/PhysRevB.83.104512} {\bibfield
  {journal} {\bibinfo  {journal} {Phys. Rev. B}\ }\textbf {\bibinfo {volume}
  {83}},\ \bibinfo {pages} {104512} (\bibinfo {year} {2011})}\BibitemShut
  {NoStop}%
\bibitem [{\citenamefont {Spa\l{}ek}\ \emph {et~al.}(2017)\citenamefont
  {Spa\l{}ek}, \citenamefont {Zegrodnik},\ and\ \citenamefont
  {Kaczmarczyk}}]{Zegrodnik2}%
  \BibitemOpen
  \bibfield  {author} {\bibinfo {author} {\bibfnamefont {J.}~\bibnamefont
  {Spa\l{}ek}}, \bibinfo {author} {\bibfnamefont {M.}~\bibnamefont
  {Zegrodnik}},\ and\ \bibinfo {author} {\bibfnamefont {J.}~\bibnamefont
  {Kaczmarczyk}},\ }\bibfield  {title} {\bibinfo {title} {Universal properties
  of high-temperature superconductors from real-space pairing:
  $t\ensuremath{-}j\ensuremath{-}u$ model and its quantitative comparison with
  experiment},\ }\href {https://doi.org/10.1103/PhysRevB.95.024506} {\bibfield
  {journal} {\bibinfo  {journal} {Phys. Rev. B}\ }\textbf {\bibinfo {volume}
  {95}},\ \bibinfo {pages} {024506} (\bibinfo {year} {2017})}\BibitemShut
  {NoStop}%
\bibitem [{\citenamefont {Zegrodnik}\ and\ \citenamefont
  {Spa\l{}ek}(2017)}]{Zegrodnik1}%
  \BibitemOpen
  \bibfield  {author} {\bibinfo {author} {\bibfnamefont {M.}~\bibnamefont
  {Zegrodnik}}\ and\ \bibinfo {author} {\bibfnamefont {J.}~\bibnamefont
  {Spa\l{}ek}},\ }\bibfield  {title} {\bibinfo {title} {Universal properties of
  high-temperature superconductors from real-space pairing: Role of correlated
  hopping and intersite Coulomb interaction within the
  $t\text{\ensuremath{-}}j\text{\ensuremath{-}}u$ model},\ }\href
  {https://doi.org/10.1103/PhysRevB.96.054511} {\bibfield  {journal} {\bibinfo
  {journal} {Phys. Rev. B}\ }\textbf {\bibinfo {volume} {96}},\ \bibinfo
  {pages} {054511} (\bibinfo {year} {2017})}\BibitemShut {NoStop}%
\bibitem [{\citenamefont {Fabrizio}\ \emph {et~al.}(1999)\citenamefont
  {Fabrizio}, \citenamefont {Gogolin},\ and\ \citenamefont
  {Nersesyan}}]{Fabrizio_1999}%
  \BibitemOpen
  \bibfield  {author} {\bibinfo {author} {\bibfnamefont {M.}~\bibnamefont
  {Fabrizio}}, \bibinfo {author} {\bibfnamefont {A.~O.}\ \bibnamefont
  {Gogolin}},\ and\ \bibinfo {author} {\bibfnamefont {A.~A.}\ \bibnamefont
  {Nersesyan}},\ }\bibfield  {title} {\bibinfo {title} {From band insulator to
  mott insulator in one dimension},\ }\href
  {https://doi.org/10.1103/PhysRevLett.83.2014} {\bibfield  {journal} {\bibinfo
   {journal} {Phys. Rev. Lett.}\ }\textbf {\bibinfo {volume} {83}},\ \bibinfo
  {pages} {2014} (\bibinfo {year} {1999})}\BibitemShut {NoStop}%
\bibitem [{\citenamefont {Torio}\ \emph {et~al.}(2001)\citenamefont {Torio},
  \citenamefont {Aligia},\ and\ \citenamefont {Ceccatto}}]{Torio_2001}%
  \BibitemOpen
  \bibfield  {author} {\bibinfo {author} {\bibfnamefont {M.~E.}\ \bibnamefont
  {Torio}}, \bibinfo {author} {\bibfnamefont {A.~A.}\ \bibnamefont {Aligia}},\
  and\ \bibinfo {author} {\bibfnamefont {H.~A.}\ \bibnamefont {Ceccatto}},\
  }\bibfield  {title} {\bibinfo {title} {Phase diagram of the Hubbard chain
  with two atoms per cell},\ }\href
  {https://doi.org/10.1103/PhysRevB.64.121105} {\bibfield  {journal} {\bibinfo
  {journal} {Phys. Rev. B}\ }\textbf {\bibinfo {volume} {64}},\ \bibinfo
  {pages} {121105} (\bibinfo {year} {2001})}\BibitemShut {NoStop}%
\bibitem [{\citenamefont {Anusooya-Pati}\ \emph {et~al.}(2001)\citenamefont
  {Anusooya-Pati}, \citenamefont {Soos},\ and\ \citenamefont
  {Painelli}}]{Painelli_2001}%
  \BibitemOpen
  \bibfield  {author} {\bibinfo {author} {\bibfnamefont {Y.}~\bibnamefont
  {Anusooya-Pati}}, \bibinfo {author} {\bibfnamefont {Z.~G.}\ \bibnamefont
  {Soos}},\ and\ \bibinfo {author} {\bibfnamefont {A.}~\bibnamefont
  {Painelli}},\ }\bibfield  {title} {\bibinfo {title} {Symmetry crossover and
  excitation thresholds at the neutral-ionic transition of the modified Hubbard
  model},\ }\href {https://doi.org/10.1103/PhysRevB.63.205118} {\bibfield
  {journal} {\bibinfo  {journal} {Phys. Rev. B}\ }\textbf {\bibinfo {volume}
  {63}},\ \bibinfo {pages} {205118} (\bibinfo {year} {2001})}\BibitemShut
  {NoStop}%
\bibitem [{\citenamefont {Wilkens}\ and\ \citenamefont
  {Martin}(2001)}]{Martin_2001}%
  \BibitemOpen
  \bibfield  {author} {\bibinfo {author} {\bibfnamefont {T.}~\bibnamefont
  {Wilkens}}\ and\ \bibinfo {author} {\bibfnamefont {R.~M.}\ \bibnamefont
  {Martin}},\ }\bibfield  {title} {\bibinfo {title} {Quantum monte carlo study
  of the one-dimensional ionic Hubbard model},\ }\href
  {https://doi.org/10.1103/PhysRevB.63.235108} {\bibfield  {journal} {\bibinfo
  {journal} {Phys. Rev. B}\ }\textbf {\bibinfo {volume} {63}},\ \bibinfo
  {pages} {235108} (\bibinfo {year} {2001})}\BibitemShut {NoStop}%
\bibitem [{\citenamefont {Kampf}\ \emph {et~al.}(2003)\citenamefont {Kampf},
  \citenamefont {Sekania}, \citenamefont {Japaridze},\ and\ \citenamefont
  {Brune}}]{Kampf_2003}%
  \BibitemOpen
  \bibfield  {author} {\bibinfo {author} {\bibfnamefont {A.~P.}\ \bibnamefont
  {Kampf}}, \bibinfo {author} {\bibfnamefont {M.}~\bibnamefont {Sekania}},
  \bibinfo {author} {\bibfnamefont {G.~I.}\ \bibnamefont {Japaridze}},\ and\
  \bibinfo {author} {\bibfnamefont {P.}~\bibnamefont {Brune}},\ }\bibfield
  {title} {\bibinfo {title} {Nature of the insulating phases in the half-filled
  ionic Hubbard model},\ }\href {https://doi.org/10.1088/0953-8984/15/34/319}
  {\bibfield  {journal} {\bibinfo  {journal} {Journal of Physics: Condensed
  Matter}\ }\textbf {\bibinfo {volume} {15}},\ \bibinfo {pages} {5895}
  (\bibinfo {year} {2003})}\BibitemShut {NoStop}%
\bibitem [{\citenamefont {Manmana}\ \emph {et~al.}(2004)\citenamefont
  {Manmana}, \citenamefont {Meden}, \citenamefont {Noack},\ and\ \citenamefont
  {Sch\"onhammer}}]{Manmana_2004}%
  \BibitemOpen
  \bibfield  {author} {\bibinfo {author} {\bibfnamefont {S.~R.}\ \bibnamefont
  {Manmana}}, \bibinfo {author} {\bibfnamefont {V.}~\bibnamefont {Meden}},
  \bibinfo {author} {\bibfnamefont {R.~M.}\ \bibnamefont {Noack}},\ and\
  \bibinfo {author} {\bibfnamefont {K.}~\bibnamefont {Sch\"onhammer}},\
  }\bibfield  {title} {\bibinfo {title} {Quantum critical behavior of the
  one-dimensional ionic Hubbard model},\ }\href
  {https://doi.org/10.1103/PhysRevB.70.155115} {\bibfield  {journal} {\bibinfo
  {journal} {Phys. Rev. B}\ }\textbf {\bibinfo {volume} {70}},\ \bibinfo
  {pages} {155115} (\bibinfo {year} {2004})}\BibitemShut {NoStop}%
\bibitem [{\citenamefont {Jabben}\ \emph {et~al.}(2005)\citenamefont {Jabben},
  \citenamefont {Grewe},\ and\ \citenamefont {Anders}}]{Jabben_2005}%
  \BibitemOpen
  \bibfield  {author} {\bibinfo {author} {\bibfnamefont {T.}~\bibnamefont
  {Jabben}}, \bibinfo {author} {\bibfnamefont {N.}~\bibnamefont {Grewe}},\ and\
  \bibinfo {author} {\bibfnamefont {F.~B.}\ \bibnamefont {Anders}},\ }\bibfield
   {title} {\bibinfo {title} {Charge gaps and quasiparticle bands of the ionic
  Hubbard model},\ }\href {https://doi.org/10.1140/epjb/e2005-00098-2}
  {\bibfield  {journal} {\bibinfo  {journal} {The European Physical Journal B -
  Condensed Matter and Complex Systems}\ }\textbf {\bibinfo {volume} {44}},\
  \bibinfo {pages} {47} (\bibinfo {year} {2005})}\BibitemShut {NoStop}%
\bibitem [{\citenamefont {Garg}\ \emph {et~al.}(2006)\citenamefont {Garg},
  \citenamefont {Krishnamurthy},\ and\ \citenamefont {Randeria}}]{Garg_2006}%
  \BibitemOpen
  \bibfield  {author} {\bibinfo {author} {\bibfnamefont {A.}~\bibnamefont
  {Garg}}, \bibinfo {author} {\bibfnamefont {H.~R.}\ \bibnamefont
  {Krishnamurthy}},\ and\ \bibinfo {author} {\bibfnamefont {M.}~\bibnamefont
  {Randeria}},\ }\bibfield  {title} {\bibinfo {title} {Can correlations drive a
  band insulator metallic?},\ }\href
  {https://doi.org/10.1103/PhysRevLett.97.046403} {\bibfield  {journal}
  {\bibinfo  {journal} {Phys. Rev. Lett.}\ }\textbf {\bibinfo {volume} {97}},\
  \bibinfo {pages} {046403} (\bibinfo {year} {2006})}\BibitemShut {NoStop}%
\bibitem [{\citenamefont {Bouadim}\ \emph {et~al.}(2007)\citenamefont
  {Bouadim}, \citenamefont {Paris}, \citenamefont {H\'ebert}, \citenamefont
  {Batrouni},\ and\ \citenamefont {Scalettar}}]{Scalettar_2007}%
  \BibitemOpen
  \bibfield  {author} {\bibinfo {author} {\bibfnamefont {K.}~\bibnamefont
  {Bouadim}}, \bibinfo {author} {\bibfnamefont {N.}~\bibnamefont {Paris}},
  \bibinfo {author} {\bibfnamefont {F.}~\bibnamefont {H\'ebert}}, \bibinfo
  {author} {\bibfnamefont {G.~G.}\ \bibnamefont {Batrouni}},\ and\ \bibinfo
  {author} {\bibfnamefont {R.~T.}\ \bibnamefont {Scalettar}},\ }\bibfield
  {title} {\bibinfo {title} {Metallic phase in the two-dimensional ionic
  Hubbard model},\ }\href {https://doi.org/10.1103/PhysRevB.76.085112}
  {\bibfield  {journal} {\bibinfo  {journal} {Phys. Rev. B}\ }\textbf {\bibinfo
  {volume} {76}},\ \bibinfo {pages} {085112} (\bibinfo {year}
  {2007})}\BibitemShut {NoStop}%
\bibitem [{\citenamefont {Kancharla}\ and\ \citenamefont
  {Dagotto}(2007)}]{Dagotto_2007}%
  \BibitemOpen
  \bibfield  {author} {\bibinfo {author} {\bibfnamefont {S.~S.}\ \bibnamefont
  {Kancharla}}\ and\ \bibinfo {author} {\bibfnamefont {E.}~\bibnamefont
  {Dagotto}},\ }\bibfield  {title} {\bibinfo {title} {Correlated insulated
  phase suggests bond order between band and mott insulators in two
  dimensions},\ }\href {https://doi.org/10.1103/PhysRevLett.98.016402}
  {\bibfield  {journal} {\bibinfo  {journal} {Phys. Rev. Lett.}\ }\textbf
  {\bibinfo {volume} {98}},\ \bibinfo {pages} {016402} (\bibinfo {year}
  {2007})}\BibitemShut {NoStop}%
\bibitem [{\citenamefont {Craco}\ \emph {et~al.}(2008)\citenamefont {Craco},
  \citenamefont {Lombardo}, \citenamefont {Hayn}, \citenamefont {Japaridze},\
  and\ \citenamefont {M\"uller-Hartmann}}]{Craco_2008}%
  \BibitemOpen
  \bibfield  {author} {\bibinfo {author} {\bibfnamefont {L.}~\bibnamefont
  {Craco}}, \bibinfo {author} {\bibfnamefont {P.}~\bibnamefont {Lombardo}},
  \bibinfo {author} {\bibfnamefont {R.}~\bibnamefont {Hayn}}, \bibinfo {author}
  {\bibfnamefont {G.~I.}\ \bibnamefont {Japaridze}},\ and\ \bibinfo {author}
  {\bibfnamefont {E.}~\bibnamefont {M\"uller-Hartmann}},\ }\bibfield  {title}
  {\bibinfo {title} {Electronic phase transitions in the half-filled ionic
  Hubbard model},\ }\href {https://doi.org/10.1103/PhysRevB.78.075121}
  {\bibfield  {journal} {\bibinfo  {journal} {Phys. Rev. B}\ }\textbf {\bibinfo
  {volume} {78}},\ \bibinfo {pages} {075121} (\bibinfo {year}
  {2008})}\BibitemShut {NoStop}%
\bibitem [{\citenamefont {Byczuk}\ \emph {et~al.}(2009)\citenamefont {Byczuk},
  \citenamefont {Sekania}, \citenamefont {Hofstetter},\ and\ \citenamefont
  {Kampf}}]{Byczuk_2009}%
  \BibitemOpen
  \bibfield  {author} {\bibinfo {author} {\bibfnamefont {K.}~\bibnamefont
  {Byczuk}}, \bibinfo {author} {\bibfnamefont {M.}~\bibnamefont {Sekania}},
  \bibinfo {author} {\bibfnamefont {W.}~\bibnamefont {Hofstetter}},\ and\
  \bibinfo {author} {\bibfnamefont {A.~P.}\ \bibnamefont {Kampf}},\ }\bibfield
  {title} {\bibinfo {title} {Insulating behavior with spin and charge order in
  the ionic Hubbard model},\ }\href
  {https://doi.org/10.1103/PhysRevB.79.121103} {\bibfield  {journal} {\bibinfo
  {journal} {Phys. Rev. B}\ }\textbf {\bibinfo {volume} {79}},\ \bibinfo
  {pages} {121103} (\bibinfo {year} {2009})}\BibitemShut {NoStop}%
\bibitem [{\citenamefont {Tincani}\ \emph {et~al.}(2009)\citenamefont
  {Tincani}, \citenamefont {Noack},\ and\ \citenamefont
  {Baeriswyl}}]{Baeriswyl_2009}%
  \BibitemOpen
  \bibfield  {author} {\bibinfo {author} {\bibfnamefont {L.}~\bibnamefont
  {Tincani}}, \bibinfo {author} {\bibfnamefont {R.~M.}\ \bibnamefont {Noack}},\
  and\ \bibinfo {author} {\bibfnamefont {D.}~\bibnamefont {Baeriswyl}},\
  }\bibfield  {title} {\bibinfo {title} {Critical properties of the
  band-insulator-to-mott-insulator transition in the strong-coupling limit of
  the ionic Hubbard model},\ }\href
  {https://doi.org/10.1103/PhysRevB.79.165109} {\bibfield  {journal} {\bibinfo
  {journal} {Phys. Rev. B}\ }\textbf {\bibinfo {volume} {79}},\ \bibinfo
  {pages} {165109} (\bibinfo {year} {2009})}\BibitemShut {NoStop}%
\bibitem [{\citenamefont {Hoang}(2010)}]{Hoang_2010}%
  \BibitemOpen
  \bibfield  {author} {\bibinfo {author} {\bibfnamefont {A.~T.}\ \bibnamefont
  {Hoang}},\ }\bibfield  {title} {\bibinfo {title} {Metal–insulator
  transitions in the half-filled ionic Hubbard model},\ }\href
  {https://doi.org/10.1088/0953-8984/22/9/095602} {\bibfield  {journal}
  {\bibinfo  {journal} {Journal of Physics: Condensed Matter}\ }\textbf
  {\bibinfo {volume} {22}},\ \bibinfo {pages} {095602} (\bibinfo {year}
  {2010})}\BibitemShut {NoStop}%
\bibitem [{\citenamefont {Torrance}\ \emph {et~al.}(1981)\citenamefont
  {Torrance}, \citenamefont {Girlando}, \citenamefont {Mayerle}, \citenamefont
  {Crowley}, \citenamefont {Lee}, \citenamefont {Batail},\ and\ \citenamefont
  {LaPlaca}}]{LaPlaca_1981}%
  \BibitemOpen
  \bibfield  {author} {\bibinfo {author} {\bibfnamefont {J.~B.}\ \bibnamefont
  {Torrance}}, \bibinfo {author} {\bibfnamefont {A.}~\bibnamefont {Girlando}},
  \bibinfo {author} {\bibfnamefont {J.~J.}\ \bibnamefont {Mayerle}}, \bibinfo
  {author} {\bibfnamefont {J.~I.}\ \bibnamefont {Crowley}}, \bibinfo {author}
  {\bibfnamefont {V.~Y.}\ \bibnamefont {Lee}}, \bibinfo {author} {\bibfnamefont
  {P.}~\bibnamefont {Batail}},\ and\ \bibinfo {author} {\bibfnamefont {S.~J.}\
  \bibnamefont {LaPlaca}},\ }\bibfield  {title} {\bibinfo {title} {Anomalous
  nature of neutral-to-ionic phase transition in
  tetrathiafulvalene-chloranil},\ }\href
  {https://doi.org/10.1103/PhysRevLett.47.1747} {\bibfield  {journal} {\bibinfo
   {journal} {Phys. Rev. Lett.}\ }\textbf {\bibinfo {volume} {47}},\ \bibinfo
  {pages} {1747} (\bibinfo {year} {1981})}\BibitemShut {NoStop}%
\bibitem [{\citenamefont {Egami}\ \emph {et~al.}(1993)\citenamefont {Egami},
  \citenamefont {Ishihara},\ and\ \citenamefont {Tachiki}}]{Egami_1993}%
  \BibitemOpen
  \bibfield  {author} {\bibinfo {author} {\bibfnamefont {T.}~\bibnamefont
  {Egami}}, \bibinfo {author} {\bibfnamefont {S.}~\bibnamefont {Ishihara}},\
  and\ \bibinfo {author} {\bibfnamefont {M.}~\bibnamefont {Tachiki}},\
  }\bibfield  {title} {\bibinfo {title} {Lattice effect of strong electron
  correlation: Implication for ferroelectricity and superconductivity},\ }\href
  {https://doi.org/10.1126/science.261.5126.1307} {\bibfield  {journal}
  {\bibinfo  {journal} {Science}\ }\textbf {\bibinfo {volume} {261}},\ \bibinfo
  {pages} {1307} (\bibinfo {year} {1993})}\BibitemShut {NoStop}%
\bibitem [{\citenamefont {Kennes}\ \emph {et~al.}(2020)\citenamefont {Kennes},
  \citenamefont {Xian}, \citenamefont {Claassen},\ and\ \citenamefont
  {Rubio}}]{Rubio_2020}%
  \BibitemOpen
  \bibfield  {author} {\bibinfo {author} {\bibfnamefont {D.~M.}\ \bibnamefont
  {Kennes}}, \bibinfo {author} {\bibfnamefont {L.}~\bibnamefont {Xian}},
  \bibinfo {author} {\bibfnamefont {M.}~\bibnamefont {Claassen}},\ and\
  \bibinfo {author} {\bibfnamefont {A.}~\bibnamefont {Rubio}},\ }\bibfield
  {title} {\bibinfo {title} {One-dimensional flat bands in twisted bilayer
  germanium selenide},\ }\href {https://doi.org/10.1038/s41467-020-14947-0}
  {\bibfield  {journal} {\bibinfo  {journal} {Nature Communications}\ }\textbf
  {\bibinfo {volume} {11}},\ \bibinfo {pages} {1124} (\bibinfo {year}
  {2020})}\BibitemShut {NoStop}%
\bibitem [{\citenamefont {Messer}\ \emph {et~al.}(2015)\citenamefont {Messer},
  \citenamefont {Desbuquois}, \citenamefont {Uehlinger}, \citenamefont {Jotzu},
  \citenamefont {Huber}, \citenamefont {Greif},\ and\ \citenamefont
  {Esslinger}}]{Esslinger_2015}%
  \BibitemOpen
  \bibfield  {author} {\bibinfo {author} {\bibfnamefont {M.}~\bibnamefont
  {Messer}}, \bibinfo {author} {\bibfnamefont {R.}~\bibnamefont {Desbuquois}},
  \bibinfo {author} {\bibfnamefont {T.}~\bibnamefont {Uehlinger}}, \bibinfo
  {author} {\bibfnamefont {G.}~\bibnamefont {Jotzu}}, \bibinfo {author}
  {\bibfnamefont {S.}~\bibnamefont {Huber}}, \bibinfo {author} {\bibfnamefont
  {D.}~\bibnamefont {Greif}},\ and\ \bibinfo {author} {\bibfnamefont
  {T.}~\bibnamefont {Esslinger}},\ }\bibfield  {title} {\bibinfo {title}
  {Exploring competing density order in the ionic Hubbard model with ultracold
  fermions},\ }\href {https://doi.org/10.1103/PhysRevLett.115.115303}
  {\bibfield  {journal} {\bibinfo  {journal} {Phys. Rev. Lett.}\ }\textbf
  {\bibinfo {volume} {115}},\ \bibinfo {pages} {115303} (\bibinfo {year}
  {2015})}\BibitemShut {NoStop}%
\bibitem [{\citenamefont {Loida}\ \emph {et~al.}(2017)\citenamefont {Loida},
  \citenamefont {Bernier}, \citenamefont {Citro}, \citenamefont {Orignac},\
  and\ \citenamefont {Kollath}}]{Kollath_2017}%
  \BibitemOpen
  \bibfield  {author} {\bibinfo {author} {\bibfnamefont {K.}~\bibnamefont
  {Loida}}, \bibinfo {author} {\bibfnamefont {J.-S.}\ \bibnamefont {Bernier}},
  \bibinfo {author} {\bibfnamefont {R.}~\bibnamefont {Citro}}, \bibinfo
  {author} {\bibfnamefont {E.}~\bibnamefont {Orignac}},\ and\ \bibinfo {author}
  {\bibfnamefont {C.}~\bibnamefont {Kollath}},\ }\bibfield  {title} {\bibinfo
  {title} {Probing the bond order wave phase transitions of the ionic Hubbard
  model by superlattice modulation spectroscopy},\ }\href
  {https://doi.org/10.1103/PhysRevLett.119.230403} {\bibfield  {journal}
  {\bibinfo  {journal} {Phys. Rev. Lett.}\ }\textbf {\bibinfo {volume} {119}},\
  \bibinfo {pages} {230403} (\bibinfo {year} {2017})}\BibitemShut {NoStop}%
\bibitem [{\citenamefont {Viebahn}\ \emph {et~al.}(2024)\citenamefont
  {Viebahn}, \citenamefont {Walter}, \citenamefont {Bertok}, \citenamefont
  {Zhu}, \citenamefont {G\"achter}, \citenamefont {Aligia}, \citenamefont
  {Heidrich-Meisner},\ and\ \citenamefont {Esslinger}}]{Esslinger_2024}%
  \BibitemOpen
  \bibfield  {author} {\bibinfo {author} {\bibfnamefont {K.}~\bibnamefont
  {Viebahn}}, \bibinfo {author} {\bibfnamefont {A.-S.}\ \bibnamefont {Walter}},
  \bibinfo {author} {\bibfnamefont {E.}~\bibnamefont {Bertok}}, \bibinfo
  {author} {\bibfnamefont {Z.}~\bibnamefont {Zhu}}, \bibinfo {author}
  {\bibfnamefont {M.}~\bibnamefont {G\"achter}}, \bibinfo {author}
  {\bibfnamefont {A.~A.}\ \bibnamefont {Aligia}}, \bibinfo {author}
  {\bibfnamefont {F.}~\bibnamefont {Heidrich-Meisner}},\ and\ \bibinfo {author}
  {\bibfnamefont {T.}~\bibnamefont {Esslinger}},\ }\bibfield  {title} {\bibinfo
  {title} {Interactions enable Thouless pumping in a nonsliding lattice},\
  }\href {https://doi.org/10.1103/PhysRevX.14.021049} {\bibfield  {journal}
  {\bibinfo  {journal} {Phys. Rev. X}\ }\textbf {\bibinfo {volume} {14}},\
  \bibinfo {pages} {021049} (\bibinfo {year} {2024})}\BibitemShut {NoStop}%
\bibitem [{\citenamefont {Roura-Bas}\ and\ \citenamefont
  {Aligia}(2023)}]{Aligia_2023}%
  \BibitemOpen
  \bibfield  {author} {\bibinfo {author} {\bibfnamefont {P.}~\bibnamefont
  {Roura-Bas}}\ and\ \bibinfo {author} {\bibfnamefont {A.~A.}\ \bibnamefont
  {Aligia}},\ }\bibfield  {title} {\bibinfo {title} {Phase diagram of the ionic
  Hubbard model with density-dependent hopping},\ }\href
  {https://doi.org/10.1103/PhysRevB.108.115132} {\bibfield  {journal} {\bibinfo
   {journal} {Phys. Rev. B}\ }\textbf {\bibinfo {volume} {108}},\ \bibinfo
  {pages} {115132} (\bibinfo {year} {2023})}\BibitemShut {NoStop}%
\bibitem [{\citenamefont {De~Marco}\ \emph {et~al.}(2022)\citenamefont
  {De~Marco}, \citenamefont {Tolle}, \citenamefont {Halati}, \citenamefont
  {Sheikhan}, \citenamefont {L\"auchli},\ and\ \citenamefont
  {Kollath}}]{Kollath_2022}%
  \BibitemOpen
  \bibfield  {author} {\bibinfo {author} {\bibfnamefont {J.}~\bibnamefont
  {De~Marco}}, \bibinfo {author} {\bibfnamefont {L.}~\bibnamefont {Tolle}},
  \bibinfo {author} {\bibfnamefont {C.-M.}\ \bibnamefont {Halati}}, \bibinfo
  {author} {\bibfnamefont {A.}~\bibnamefont {Sheikhan}}, \bibinfo {author}
  {\bibfnamefont {A.~M.}\ \bibnamefont {L\"auchli}},\ and\ \bibinfo {author}
  {\bibfnamefont {C.}~\bibnamefont {Kollath}},\ }\bibfield  {title} {\bibinfo
  {title} {Level statistics of the one-dimensional ionic Hubbard model},\
  }\href {https://doi.org/10.1103/PhysRevResearch.4.033119} {\bibfield
  {journal} {\bibinfo  {journal} {Phys. Rev. Res.}\ }\textbf {\bibinfo {volume}
  {4}},\ \bibinfo {pages} {033119} (\bibinfo {year} {2022})}\BibitemShut
  {NoStop}%
\bibitem [{\citenamefont {Chattopadhyay}\ \emph {et~al.}(2019)\citenamefont
  {Chattopadhyay}, \citenamefont {Bag}, \citenamefont {Krishnamurthy},\ and\
  \citenamefont {Garg}}]{Garg_2019}%
  \BibitemOpen
  \bibfield  {author} {\bibinfo {author} {\bibfnamefont {A.}~\bibnamefont
  {Chattopadhyay}}, \bibinfo {author} {\bibfnamefont {S.}~\bibnamefont {Bag}},
  \bibinfo {author} {\bibfnamefont {H.~R.}\ \bibnamefont {Krishnamurthy}},\
  and\ \bibinfo {author} {\bibfnamefont {A.}~\bibnamefont {Garg}},\ }\bibfield
  {title} {\bibinfo {title} {Phase diagram of the half-filled ionic Hubbard
  model in the limit of strong correlations},\ }\href
  {https://doi.org/10.1103/PhysRevB.99.155127} {\bibfield  {journal} {\bibinfo
  {journal} {Phys. Rev. B}\ }\textbf {\bibinfo {volume} {99}},\ \bibinfo
  {pages} {155127} (\bibinfo {year} {2019})}\BibitemShut {NoStop}%
\bibitem [{\citenamefont {Wang}\ \emph {et~al.}(2022)\citenamefont {Wang},
  \citenamefont {Wang},\ and\ \citenamefont {Wang}}]{Wang_2022}%
  \BibitemOpen
  \bibfield  {author} {\bibinfo {author} {\bibfnamefont {S.-Y.}\ \bibnamefont
  {Wang}}, \bibinfo {author} {\bibfnamefont {D.}~\bibnamefont {Wang}},\ and\
  \bibinfo {author} {\bibfnamefont {Q.-H.}\ \bibnamefont {Wang}},\ }\bibfield
  {title} {\bibinfo {title} {Transition from band insulator to mott insulator
  and formation of local moment in the half-filled ionic $\mathrm{SU}(n)$
  Hubbard model},\ }\href {https://doi.org/10.1103/PhysRevB.106.245113}
  {\bibfield  {journal} {\bibinfo  {journal} {Phys. Rev. B}\ }\textbf {\bibinfo
  {volume} {106}},\ \bibinfo {pages} {245113} (\bibinfo {year}
  {2022})}\BibitemShut {NoStop}%
\bibitem [{\citenamefont {Lanat\`a}\ \emph {et~al.}(2008)\citenamefont
  {Lanat\`a}, \citenamefont {Barone},\ and\ \citenamefont
  {Fabrizio}}]{Lanata2008}%
  \BibitemOpen
  \bibfield  {author} {\bibinfo {author} {\bibfnamefont {N.}~\bibnamefont
  {Lanat\`a}}, \bibinfo {author} {\bibfnamefont {P.}~\bibnamefont {Barone}},\
  and\ \bibinfo {author} {\bibfnamefont {M.}~\bibnamefont {Fabrizio}},\
  }\bibfield  {title} {\bibinfo {title} {Fermi-surface evolution across the
  magnetic phase transition in the kondo lattice model},\ }\href
  {https://doi.org/10.1103/PhysRevB.78.155127} {\bibfield  {journal} {\bibinfo
  {journal} {Phys. Rev. B}\ }\textbf {\bibinfo {volume} {78}},\ \bibinfo
  {pages} {155127} (\bibinfo {year} {2008})}\BibitemShut {NoStop}%
\bibitem [{\citenamefont {Fabrizio}(2007)}]{Fabrizio2007}%
  \BibitemOpen
  \bibfield  {author} {\bibinfo {author} {\bibfnamefont {M.}~\bibnamefont
  {Fabrizio}},\ }\bibfield  {title} {\bibinfo {title} {Gutzwiller description
  of non-magnetic mott insulators: Dimer lattice model},\ }\href
  {https://doi.org/10.1103/PhysRevB.76.165110} {\bibfield  {journal} {\bibinfo
  {journal} {Phys. Rev. B}\ }\textbf {\bibinfo {volume} {76}},\ \bibinfo
  {pages} {165110} (\bibinfo {year} {2007})}\BibitemShut {NoStop}%
\bibitem [{\citenamefont {Fabrizio}(2013)}]{Fabrizio2012}%
  \BibitemOpen
  \bibfield  {author} {\bibinfo {author} {\bibfnamefont {M.}~\bibnamefont
  {Fabrizio}},\ }\bibfield  {title} {\bibinfo {title} {The out-of-equilibrium
  time-dependent Gutzwiller approximation},\ }in\ \href@noop {} {\emph
  {\bibinfo {booktitle} {New Materials for Thermoelectric Applications: Theory
  and Experiment}}},\ \bibinfo {editor} {edited by\ \bibinfo {editor}
  {\bibfnamefont {V.}~\bibnamefont {Zlatic}}\ and\ \bibinfo {editor}
  {\bibfnamefont {A.}~\bibnamefont {Hewson}}}\ (\bibinfo  {publisher} {Springer
  Netherlands},\ \bibinfo {address} {Dordrecht},\ \bibinfo {year} {2013})\ pp.\
  \bibinfo {pages} {247--273}\BibitemShut {NoStop}%
\bibitem [{\citenamefont {Schir\'o}\ and\ \citenamefont
  {Fabrizio}(2010)}]{Schiro2010}%
  \BibitemOpen
  \bibfield  {author} {\bibinfo {author} {\bibfnamefont {M.}~\bibnamefont
  {Schir\'o}}\ and\ \bibinfo {author} {\bibfnamefont {M.}~\bibnamefont
  {Fabrizio}},\ }\bibfield  {title} {\bibinfo {title} {Time-dependent mean
  field theory for quench dynamics in correlated electron systems},\ }\href
  {https://doi.org/10.1103/PhysRevLett.105.076401} {\bibfield  {journal}
  {\bibinfo  {journal} {Phys. Rev. Lett.}\ }\textbf {\bibinfo {volume} {105}},\
  \bibinfo {pages} {076401} (\bibinfo {year} {2010})}\BibitemShut {NoStop}%
\bibitem [{\citenamefont {Wysoki\ifmmode~\acute{n}\else \'{n}\fi{}ski}\ and\
  \citenamefont {Fabrizio}(2017{\natexlab{a}})}]{Wysokinski2016}%
  \BibitemOpen
  \bibfield  {author} {\bibinfo {author} {\bibfnamefont {M.~M.}\ \bibnamefont
  {Wysoki\ifmmode~\acute{n}\else \'{n}\fi{}ski}}\ and\ \bibinfo {author}
  {\bibfnamefont {M.}~\bibnamefont {Fabrizio}},\ }\bibfield  {title} {\bibinfo
  {title} {Mott physics beyond the Brinkman-Rice scenario},\ }\href
  {https://doi.org/10.1103/PhysRevB.95.161106} {\bibfield  {journal} {\bibinfo
  {journal} {Phys. Rev. B}\ }\textbf {\bibinfo {volume} {95}},\ \bibinfo
  {pages} {161106} (\bibinfo {year} {2017}{\natexlab{a}})}\BibitemShut
  {NoStop}%
\bibitem [{\citenamefont {Wysoki\ifmmode~\acute{n}\else \'{n}\fi{}ski}\ and\
  \citenamefont {Fabrizio}(2017{\natexlab{b}})}]{Wysokinski2017}%
  \BibitemOpen
  \bibfield  {author} {\bibinfo {author} {\bibfnamefont {M.~M.}\ \bibnamefont
  {Wysoki\ifmmode~\acute{n}\else \'{n}\fi{}ski}}\ and\ \bibinfo {author}
  {\bibfnamefont {M.}~\bibnamefont {Fabrizio}},\ }\bibfield  {title} {\bibinfo
  {title} {Interplay of charge and spin dynamics after an interaction quench in
  the Hubbard model},\ }\href {https://doi.org/10.1103/PhysRevB.96.201115}
  {\bibfield  {journal} {\bibinfo  {journal} {Phys. Rev. B}\ }\textbf {\bibinfo
  {volume} {96}},\ \bibinfo {pages} {201115} (\bibinfo {year}
  {2017}{\natexlab{b}})}\BibitemShut {NoStop}%
\end{thebibliography}
\end{document}